\documentclass[reprint,superscriptaddress,aps,floatfix,prb,amsmath,amssymb]{revtex4-2}
\usepackage[utf8]{inputenc}
\usepackage{bm}
\usepackage{graphicx}
\usepackage{hyperref}
\usepackage{color}
\usepackage{nicefrac}
\usepackage{booktabs}
\usepackage{natbib}
\usepackage{amsmath}
\usepackage{xcolor}

\begin{document}

\title{Finite-size effects on the ferroelectricity in  rhombohedral HfO$_2$}

\author{Francesco Delodovici}
\affiliation{Université Paris-Saclay, CentraleSupélec, CNRS, Laboratoire SPMS, 91190, Gif-sur-Yvette, France}
\affiliation{Consiglio Nazionale delle Ricerche CNR-SPIN, c/o Universit\`a degli Studi “G. D’Annunzio”, I-66100 Chieti, Italy}

\author{Paolo Barone}
\affiliation{Consiglio Nazionale delle Ricerche CNR-SPIN, Area della Ricerca di Tor Vergata, Via del Fosso del Cavaliere, 100, I-00133 Rome, Italy }

\author{Silvia Picozzi}
\affiliation{Consiglio Nazionale delle Ricerche CNR-SPIN, c/o Universit\`a degli Studi “G. D’Annunzio”, I-66100 Chieti, Italy}

\begin{abstract}
In this work we analyze the finite-size effects on the structural properties and on the polarization of the rhombohedral phase of HfO$_2$ subjected to a biaxial compressive strain. We show how the presence of surface charges affects the polarization, leading to a strong reduction with respect to its bulk value. This reduction can be ascribed to two mechanisms: {\em i}) the coupling between compressive strain and the phase-transition order parameter; {\em ii}) the changes in the ferroelectric distortion. We give two alternative explanations of this phenomenon: from an atomistic point of view, analyzing the evolution of the bond lengths, and from a symmetry-analysis point of view, considering the changes in the amplitude of the symmetry-allowed distortions, when a slab configuration is considered. These results are independent on the slab-thickness in the considered range, suggesting the absence of a critical thickness for ferroelectricity in HfO$_2$, in agreement with the proposed improper nature of hafnia ferroelectricity.

\end{abstract}

\maketitle

\section{Introduction}
\large
Ferroelectricity in hafnium oxide has been at the center of continuous interests and deep analysis in the last decade. 
The stabilization of different crystalline phases represents a particularly delicate issue. In closer detail,
the detected polar phases \cite{Schroder_2011,Lee_2020_SCIENCE,Noheda_2018,Barabash_2017} of the pristine oxide \cite{Xu_2021} are only metastable in bulk configurations at standard thermodynamical conditions, its ground state rather being a non-polar monoclininc phase \cite{Huan_2014}.
However, the polar phases become thermodynamically competitive with the monoclinic phase when hafnia is realized in the form of films \cite{Schroder_2011}. In this respect, a variety of different factors play a role, such as
mechanical strain induced by the substrate \cite{Batra_2017,Estandia_2019,Estandia_2020,Liu_2019}, doping  \cite{Batra_2017,Schroder_2018}, 
defects \cite{Chouprik_2021,hoffmann_2015,Nukala_2021}, 
grains size and  films thickness  \cite{Polakowski_2015}. 
Despite many open issues, it is anyway beyond question that its compatibility with Si-based semiconductors \cite{Chernikova_2016}, along with the robustness of the polarization hysteresis at the nanoscale level \cite{Yurchuk_2013,Dawber_2005} and its simple chemistry, make HfO$_2$ extremely appealing for  ferroelectric-based technological applications, when compared to ordinary ferroelectric perovskite--like films \cite{Muller_2015}.

From the theoretical modelling point of view, an extremely  interesting topic is represented by the investigation of how the order parameter describing ferroelectricity couples to  surfaces and interfaces \cite{Eliseev_215,Meyer_2001}.
In closer detail, it is particularly interesting to determine whether a minimal thickness for the development of ferroelectric instabilities in the film exists or not.
The presence of such a critical thickness is generally related to the screening of  surfaces charges associated with the film spontaneous polarization via depolarizing fields \cite{Meyer_2001,junquera_2003,Batra_1973}.
In proper ferroelectrics, the occurrence of  surfaces can have dramatic consequences on the stability of the polar phase for films thinner than the critical thickness, which in the case of prototypical BaTiO$_3$ is on the nanometer order  under short-circuit condition \cite{Kim_2005,junquera_2003}. 
Ferroelectrics where the onset of polarization emerges as a secondary effect, being mainly governed by the coupling to a primary non-polar distortion, may be expected to present robustness against depolarizing effects. 
This characteristic makes improper ferroelectrics particularly appealing for their potentiality in technological applications.
Nonetheless, their behaviour in the ultra-thin limit remains unclear, as well as the existence of a critical thickness \cite{Sai_2009,Nordlander_2019}.

In this paper, we present a theoretical analysis, based on {\it ab-initio} simulations, of the effects of  finite-size on the stability of polar R3m phase of hafnium oxide.
We considered different slab configurations under open-boundary conditions with clean surfaces in the range of  experimentally-reachable nanometric thicknesses.
We stress that these electrostatic boundary conditions are highly unfavourable to proper ferroelectricity. %
Indeed, when polarization can be considered as the leading order parameter of the phase transition, these conditions correspond to vanishing electric displacement in the vacuum \cite{Meyer_2001}. 
Under this configuration, below a critical thickness, the intense depolarizing field experienced by the slab arising from the unscreened surface charges is strong enough to completely suppress the polar phase.
Further on, we analyze the evolution of  polarization as a function of the depth in the slab ({\em i.e.} distance from the surfaces) by means of  Born effective charges of  relaxed bulk phases.
Finally, we justify the observed effects in terms of microscopic modelling and in terms of suppression of the dominant symmetry-allowed distortions connecting the high and the low-symmetry phase.

\section{Slab configurations}

In this paragraph we report  some structural information about the simulations  and refer to appendices for further computational details. 
We orient the relaxed primitive cell according to the experimentally-detected growth direction \cite{Noheda_2018}. Thus, the [111] direction of the rhombohedral setting is parallel to the normal of the film surface: we chose this direction to coincide with the \textbf{z} axis in our simulations. 
In this configuration, the polarization is also directed along \textbf{z}.
Hereafter, we refer to calculations performed in the equivalent hexagonal setting.
The relaxed primitive vectors are: $\rm {\bf a}=(a,0,0)$, $\rm {\bf b}=(-a/2,a\sqrt{3}/2,0)$, $\rm {\bf c}=(0,0,c)$, where a$=7.134$ \AA, c=$8.738$ \AA. 
With this choice, the unit cell consists of 12 formula units or, equivalently, of 36 atoms.

Previous computational modelling works  \cite{Noheda_2018,Tsymbal_2020} show that the polarization undergoes a rapid increase, when biaxial compressive strain is applied normally to the polarization direction. 
In closer detail, the polarization of the relaxed bulk configuration at equilibrium  is approximately 0.06 $\mu$C/cm$^2$, as computed with the Berry phase method \cite{Vanderbilt_1993}, while higher values, above 25 $\mu$C/cm$^2$, are obtained when compressing \textbf{a} above 5\%.
Given this significant difference in the P value, we expect  finite size effects to be ``masked" in equilibrium conditions by the small polarization magnitude; rather, we expect 
finite size effects to be 
more noticeable (and therefore easier to investigate), upon application of compressive strain.
Therefore, we focus hereafter on the hexagonal cell subjected to a bi-axial 5\% in-plane compression. 
The bulk primitive cell under this strain condition is defined by a=6.78 \AA\, and c$\rm_{opt}$=9.66 \AA. 
The optimum c$\rm_{opt}$ is obtained by minimizing the total energy of the primitive cell with fixed (compressed) in-plane lattice constant and relaxed internal degrees of freedom describing atomic positions.
The polarization of this phase, computed with the Berry-phase method, is 22.8 $\mu$C/cm$^2$, in good agreement with values reported in literature \cite{Tsymbal_2020,Noheda_2019}

In order to build the slab configurations, the compressed unit cell is repeated along its [001] hexagonal direction and ``cut" at three different thicknesses. 
The resulting asymmetric slabs, before relaxation, are 12.88, 19.32, 38.64 \AA\, 
thick, corresponding  to 4 layers, 6 layers and 12 layers, respectively. Hereafter, we define a ``layer" as the  stoichiometric combination of Hf and O occupying one third of the primitive hexagonal cell volume.
As such, each layer consists of 4 Hf atoms and 8 O atoms.
With this definition, the 6 (12) layers slab consists of a primitive cell repeated 2 (4) times along the \textbf{c} axis.
%


\section{RESULTS}
\subsection{Atomic displacements}

\begin{figure}[]
\centerline{
\includegraphics[angle=0,width=0.52\textwidth]{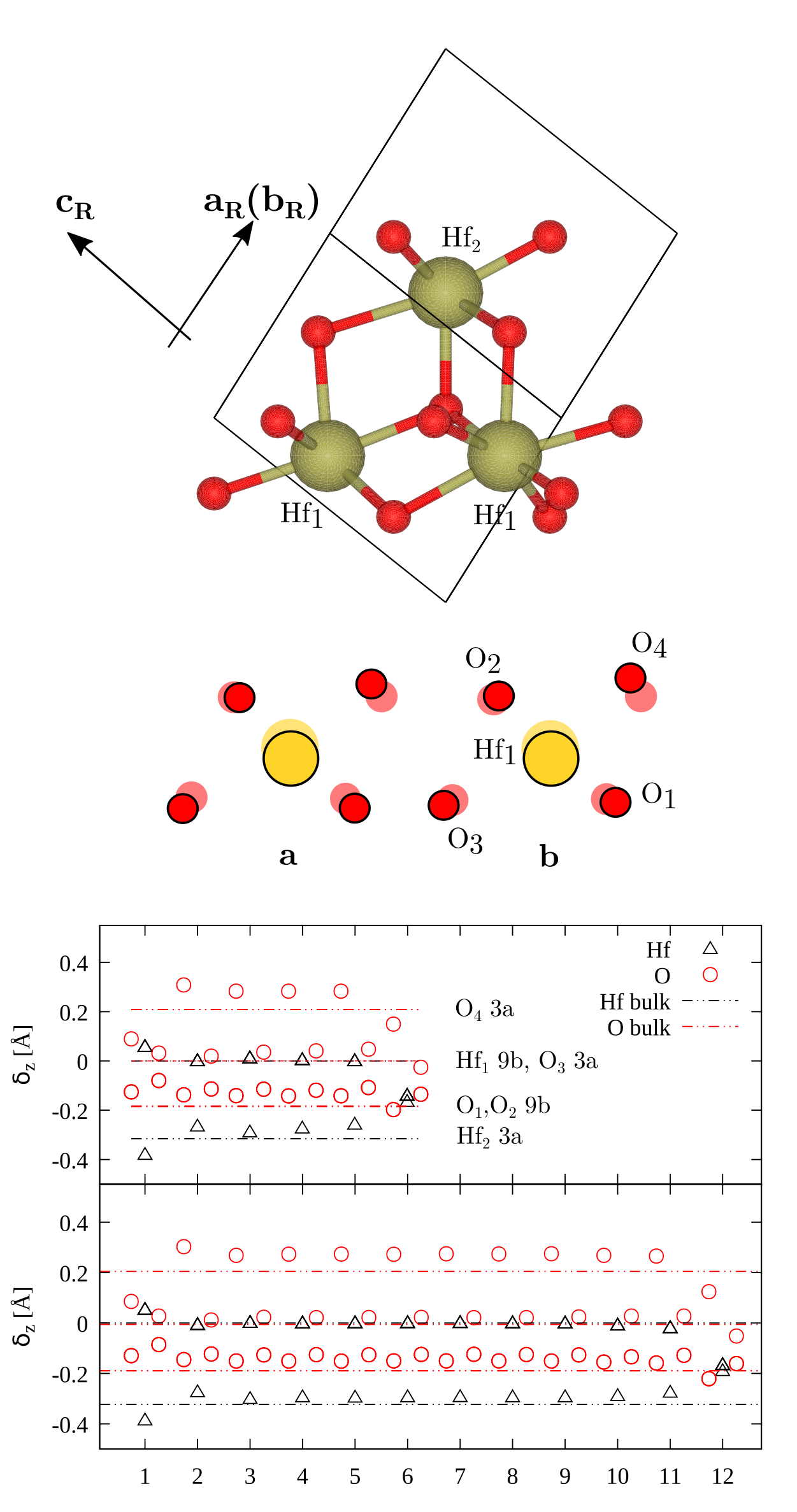}
}\caption{   Top: the primitive cell of the rhombohedral phase defined by the vectors \textbf{a$_{\rm R}$},\textbf{b$_{\rm R}$},\textbf{c$_{\rm R}$}. Hf (O) atoms shown by a yellow (red) sphere. 
    Center: panel \textbf{a} and \textbf{b} report a schematic representation of the ferroelectric distortion in bulk and slab (central layer) configurations, respectively.  The atomic positions of the paraelectric phase are represented by the shaded spheres. 
    Bottom: structural relaxation along the \textbf{c} primitive vector for the 6-layers slab (top panel) and the 12-layers slab (bottom panel). The black triangles and the red circles show, respectively, the displacements along the $z$ direction of Hf and O symmetry-inequivalent types for each layer in the relaxed slab with respect to the relaxed paraelectric R-3m. The dot-dashed lines represent the ferroelectric displacements in the bulk configurations.     
    The atoms are labelled with different Wyckoff positions and multiplicity in each panel to facilitate the comparison between the panels.
\label{displacements}}
\end{figure}

We observe the geometrical relaxation to produce two main  
effects in all the considered configurations: {\em i}) the contraction of the film thickness; {\em ii}) the relaxation of the atomic positions within the film.
To give a quantitative description of the first effect, we consider, as a marker for the contraction  experienced by the slabs, the thickness of the most central unit cell i.e. the c axis of an embedded hexagonal cell.
In fact, the central layers can be reasonably assumed to be marginally affected by the surfaces.
The comparison between the Hf-projected DOS (not shown) in these central layers with the bulk Hf-projected DOS indeed confirms such hypothesis. 
For the 4-layer-thick slab the marker is ill-defined; thus, in order to have a value comparable to those obtained for thicker slabs, we multiply the distance between the two central layers by 3.
The 12, 6 and 4-layer slab experience respectively  a contraction of $\approx$1.4\%, $\approx$1.6\%, $\approx$1.5\%.  
This contraction can be traced back to the coupling of the  strain acting along \textbf{c} with the different symmetry-allowed distortions (as will be detailed below, see section \ref{sect_modes}). 

To analyze the relaxations of  atomic positions in the film, we compare the ferroelectric distortion in the bulk phases and in the relaxed slab.
To this purpose, we consider the displacements of each atom with respect to a reference paraelectric (PE) phase: the $R\bar{3}m$.
In a bulk configuration, the polar displacements from a PE phase characterize the ferroelectric (FE) phase. 
From the same quantities, in a slab configuration, one can deduce how finite-size effects affect ferroelectricity. 
To decouple the displacements from the effects of the cell compression, we adjust the reference PE phase to have the same height of the central cell in the relaxed slab.
Fig. \ref{displacements} illustrates the vertical displacements $\rm \delta z = z_{slab} - z_{PE}$ of each atom in the HfO$_2$ slab. 
The top and bottom panel report respectively the displacement of the 6-layers and of the 12-layers slabs.
The dot-dashed lines represent the displacements characterizing the bulk phases.
The R3m bulk phase has two different Wyckoff positions for hafnium and four for oxygen, thus 6 lines appear in the panels (two oxygen lines at -0.2, in red, are almost degenerate, while a third one close to 0 overlaps with a Hf line in black) .
Each of this high-symmetry position transforms in a unique way under geometric relaxation, therefore each layer presents 4 different oxygen displacements and 2 for hafnium.
The surface layers experience the largest changes: the absolute displacements of hafnium at the bottom (first) layer change on average by 0.06 \AA\ , the oxygens closer to vacuum on average by 0.07 \AA\, whereas those toward the bulk change by 0.08 \AA\,. 
At the top layer, the hafnium displacements change by 0.16 \AA\ on average, while those of oxygens remain closer to the bulk value, changing on average by 0.03 \AA\, for O toward the bulk and of 0.04 \AA\, for O toward the vacuum. 
We stress that the four hafnium atoms at the top surface in the relaxed slab have the same $z$ coordinate.
This is analogous to what happens in the PE bulk phase.
The changes in the ferroelectric distortion rapidly converge to constant values in the central layers, with the second layer acting as an intermediate buffer. %
In the bulk region the displacements of both hafnium and oxygen atoms do not match the displacements characterizing the bulk phases: the discrepancy for oxygen reaches 0.06 \AA\, and 0.03 \AA\, for hafnium.
Based on these observations, it is reasonable to expect some changes in the electric polarization throughout the slab. 
Nonetheless, the complexity of the distortions makes these effects hard to deduce only from the evolution of the ferroelectric displacements. 
Therefore, we move to a direct analysis of the polarization throughout the slab.

\subsection{Polarization}

\begin{figure}[]
\centerline{
\includegraphics[angle=0,width=0.52\textwidth]{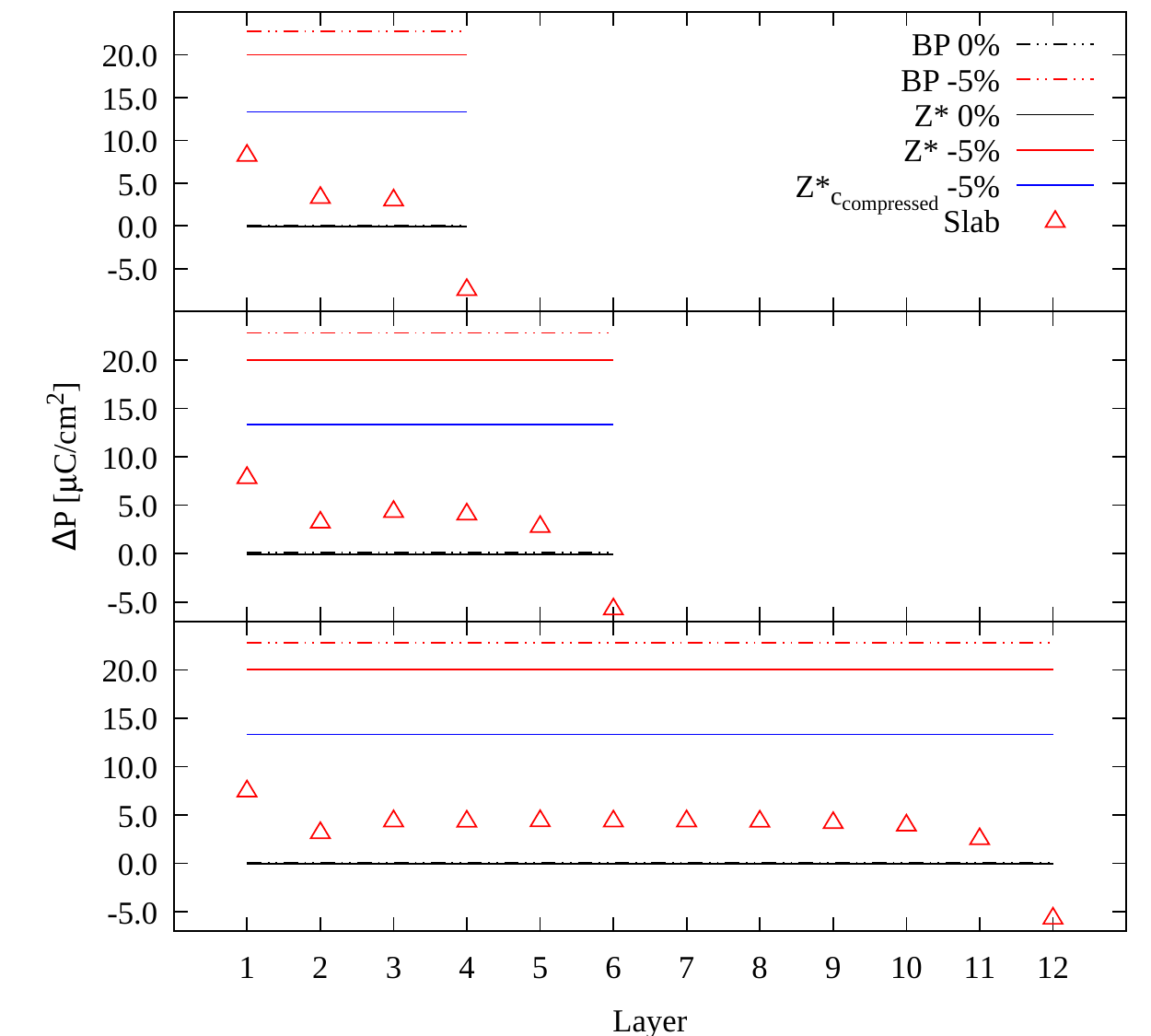}
}
\caption{ \label{polarization}
    The bulk polarization computed with different methods compared to the layer-projected value.
    Three slab thickness are reported; top: four--layers slab; center: six--layers slab; bottom: twelve--layers slab. 
    The dotted-dashed lines represent the P computed with the Berry-phase method. 
    The continuous lines report the P computed using the Born-effective charges.
    The blue line corresponds to the P of the bulk phase with compressed c.
    The triangles report the value of P per each layer.
}
\end{figure}
In order to investigate the evolution of the polarization in the slab as a function of  thickness, we employed the Born effective charges, computed for the relaxed and for the compressed bulk phases. 
We do not employ the Berry-phase method, due to the presence of metallic states confined at the top surface and associated with 
oxygen $p$-orbitals, as revealed by the computed projected density of states.
The polarization difference with respect to the centrosymmetric phase is by definition the contraction of the tensor Z$^*$ with the vector of the atomic displacements from the $R\bar{3}m$ phase, normalized to the bulk volume.
For the layer polarization, the dipole moment is normalized to the volume of the layer, hereby defined as one third of the volume of the adjusted bulk.
These simplified definitions  allow us to investigate the dependence of the dipole-moment density on the ``depth" ({\em i.e.} distance from the surfaces) within the slab.

We decouple the effects contributing to the change in polarization: the slab contraction and the local atomic distortions.
Fig. \ref{polarization} reports the evolution of the polarization as a function of the layer for three different slab thicknesses.
The polarization of different bulk configurations obtained with the BP method is reported as a comparison: 0.07 and 22.8 $\mu$C/cm$^2$ respectively for the relaxed bulk and the bulk subjected to a biaxial (along \textbf{a,b} primitive vectors) compressive (-5\%) strain. 
In this regard, it should be noticed that 
the use of effective charges to compute the polarization results in an underestimate with respect to the BP value: P computed with effective charges amounts to 0.06 and 20.03  $\mu$C/cm$^2$, respectively, for the relaxed equilibrium bulk and for the bulk subjected to a biaxial compressive (-5\%) strain.
This discrepancy  in the compressed case ({\em i.e.} almost 2.8 $\mu$C/cm$^2$) corresponds to  9\% of the BP value, which is still a reasonable margin of error for our analysis.

We stress once again that we employed the effective charges of the bulk subjected to a biaxial -5\% strain state to compute the effects of finite size on the polarization. This corresponds to assuming that the local distortions following the relaxation do not affect significantly the atomic responses to an external electric field.
In order to analyze the effects of the slab contraction, we compute the change in P arising from the compression along \textbf{c} of the relaxed R3m bulk to the value of c characterising the central hexagonal cell embedded in the 6 layers slab.
In this way, we observe that the slab contraction induces a  decrease in P by about 33\% (6.7 $\mu$C/cm$^2$) with respect to the c-relaxed bulk, as can be observed in Fig.\ref{polarization}: the polarization of the compressed bulk, 13.33 $\mu$C/cm$^2$, and the one of the c-relaxed bulk, 20.03 $\mu$C/cm$^2$, are respectively represented by the blue and the continuous red line.  Such a reduction of the polarization induced by the slab contraction can be phenomenologically traced back to the same strain-polarization coupling which is responsible for the remarkable increase of P following the 5\% bi-axial in-plane compression of the bulk configuration and consequent elongation of the cell along the out-of-plane direction. As the slab contraction causes a reduction of the strain deformation along the $c$ axis with respect to the bulk phase, the strain-polarization coupling is responsible for a corresponding reduction of P.
The introduction of finite-size causing the local rearrangement of atoms induces a further significant reduction of P.
In fact, when moving from the surfaces where the polarization assumes opposite sign and different absolute values, the polarization settles to approximately 4.5 and to 3.3 $\mu$C/cm$^2$, respectively, in the 12(6) and 4 layers slabs. 
The reduced value of P in the thinnest slab is  connected with the lack of a proper bulk-like volume in the slab, as in this configuration the central layers are just below surfaces.
In all the slabs, the polarization in the central layers is therefore reduced by more than 70\% with respect to the corresponding value in the uncompressed bulk. Nonetheless, it remains finite.
We note that the bulk Z$^*$ represent a poor approximation to the electric response at the surfaces. Therefore, the value of P at the extreme layers is likely biased and should not be commented.

It is interesting to notice that, despite the open-boundary conditions employed to simulate the slab, the depolarization field does not completely suppress the polarization, as it happens for proper-ferroelectrics slabs below their critical thickness under open-circuit conditions \cite{Meyer_2001}.
Since the polarization of the central layers remains finite as the thickness is lowered, we may infer that the critical thickness for the $R3m$ phase of HfO$_2$ films is either smaller than the thickness of the 4-layer slab, or it completely vanishes.  
We note that the extreme case of a three layer configuration still presents a finite value of P in the central layer ($\approx$ 2 $\mu$C/cm$^2$), but this can hardly be intended as a ``proper bulk--like" layer.
Therefore, the incomplete suppression of P may be an indirect sign of the improper character of ferroelectricity in this polar phase of HfO$_2$.

\section{Interpretation of polarization suppression}
\subsection{Atomistic model}

\begin{figure}[h]
\centerline{
\includegraphics[angle=-90,width=0.35\textwidth]{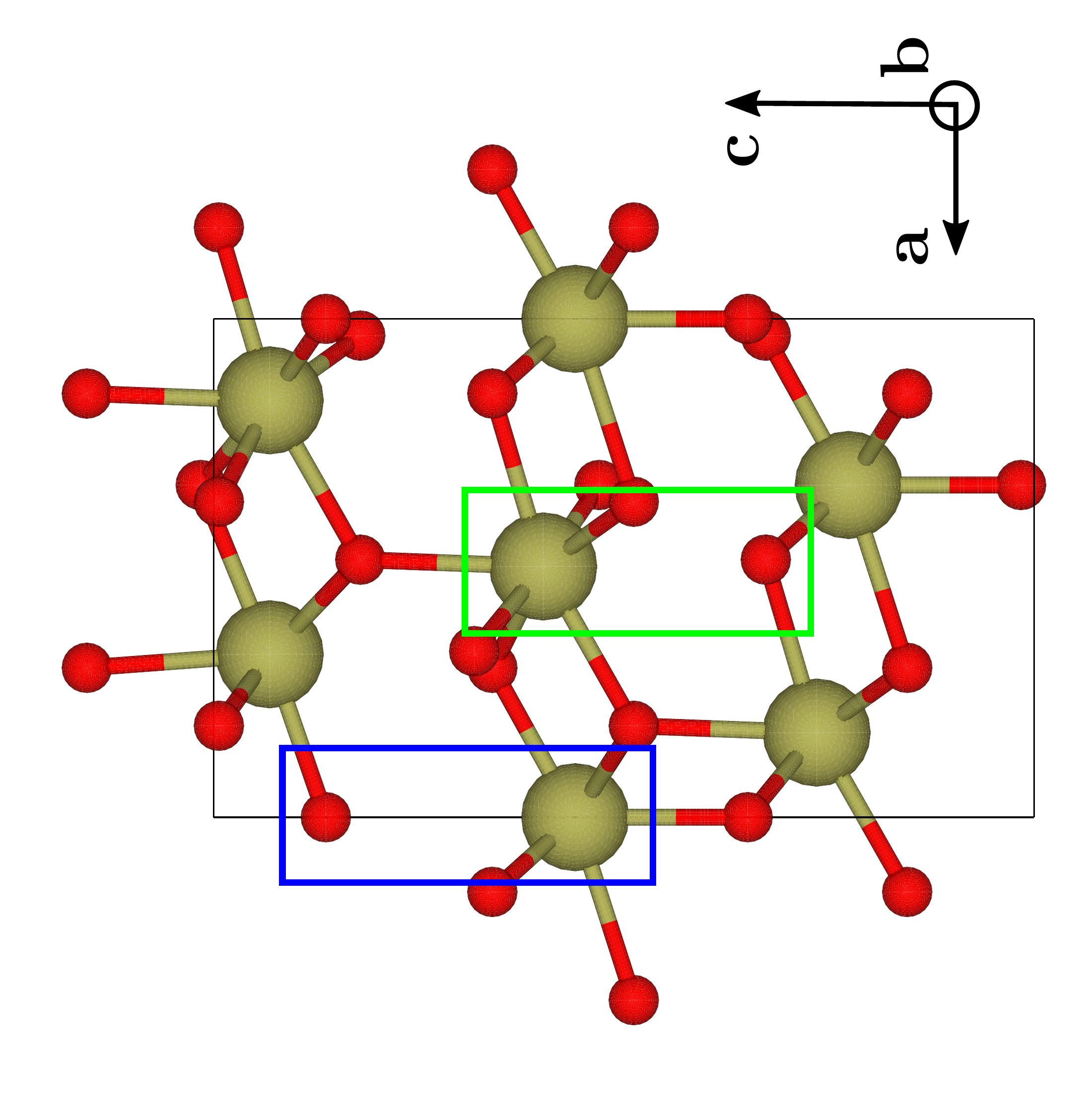}
}
\caption{ \label{prsec_bonds}
Section of the $R3m$ hexagonal unit cell. The primary and the secondary (vertical) Hf-O bonds are highlighted respectively by green and blue rectangles. In the unit cell, these bonds appear with a ratio of 3 to 1. A shorter vertical bond corresponds to both: above for the primary and below for the secondary bonds, respectively.
}
\end{figure}

\begin{figure}[]
\centerline{
\includegraphics[angle=0,width=0.5\textwidth]{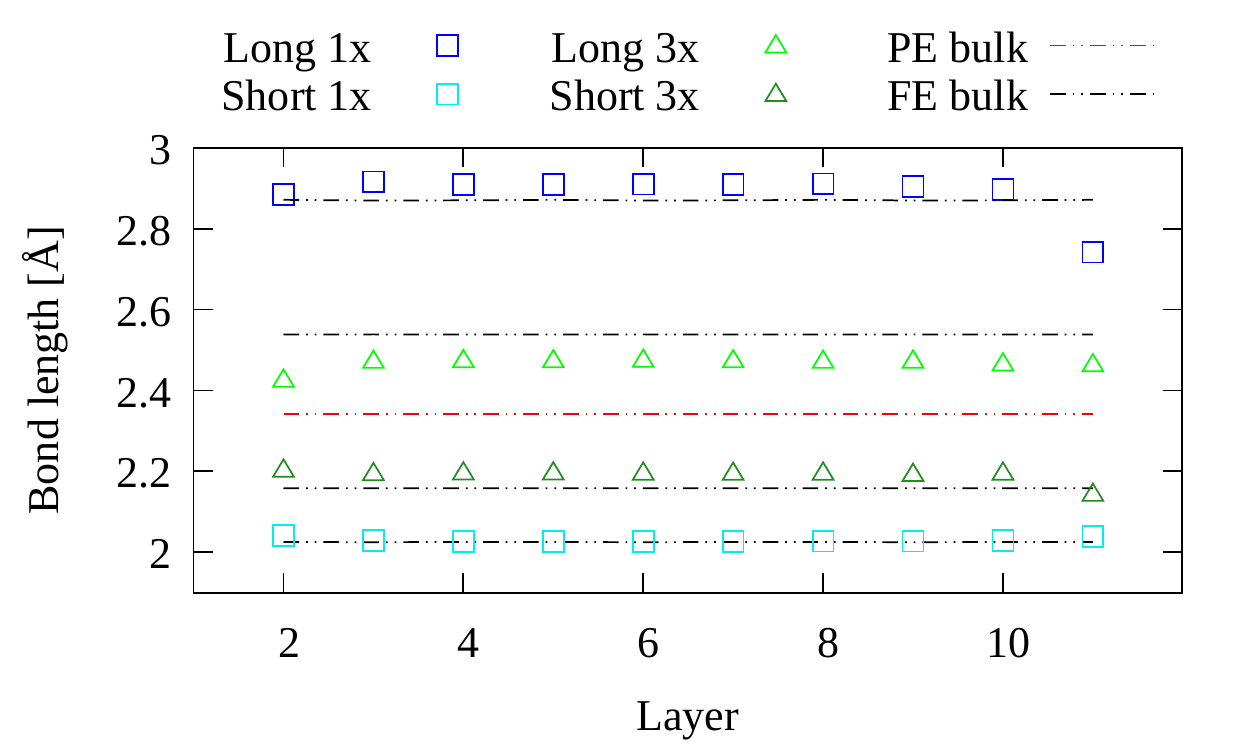}
}
\caption{ \label{bonds_length}
The trend of the long-short primary (3x) and the long-short secondary (1x) bonds in the 12-layer slab.
The dotted-dashed lines represents the corresponding bulk value. 
Each color has to be compared with the closest black line that represents the corresponding bulk bond lengths.
The red line corresponds to the value of the bonds in the $R-3m$ paraelectric phase.
The values for the two surfaces are not reported, since they reflect a different behaviour occurring at the surfaces.
}
\end{figure}

To explain the suppression of P in the central layers, we refer to the atomistic model of Zhang et al. reported in \cite{Tsymbal_2020}, where the effects of compressive strain on the $R3m$ phase are analyzed.
The authors show that the increase of P under certain strain states corresponds to the change in the contribution to P of two competing Hf-O bonds, hereafter named ``primary" and ``secondary", appearing in the unit cell with the ratio of 3 to 1.
The primary bonds involve the Hf and O with high multiplicity (9b, corresponding to Hf$_1$, O$_1$ and O$_2$ as reported in Fig. \ref{displacements}).
Instead, the secondary bonds involve the Hf and O with low multiplicity (3a, corresponding to Hf$_2$, O$_3$ and O$_4$ in Fig. \ref{displacements}). 
Fig. \ref{prsec_bonds} reports the hexagonal unit cell with a primary bond highlighted by the green rectangle and the secondary bond by the blue one.
For small strains, the contribution to the polarization of the two bonds compensate one another. 
However, above a certain strain threshold, the contribution from the 3 primary bonds grows and determines the increase of P. 
Fig. \ref{bonds_length} reports the trend of the two bonds lengths across the 12-layers slab. The secondary and the primary bonds respectively increase and decrease with respect to their bulk value.
Specifically, the primary ones approach the value of the bonds in the centro-symmetric phase.
Since the two bonds compete in determining the sign of the polarization,
 the resulting effect of such a global distortion is to reduce the polarization in the central portion of the slab.

\subsection{Suppression of the main distortion modes }
\label{sect_modes}
\begin{figure}[]
\centerline{
\includegraphics[angle=0,width=0.52\textwidth]{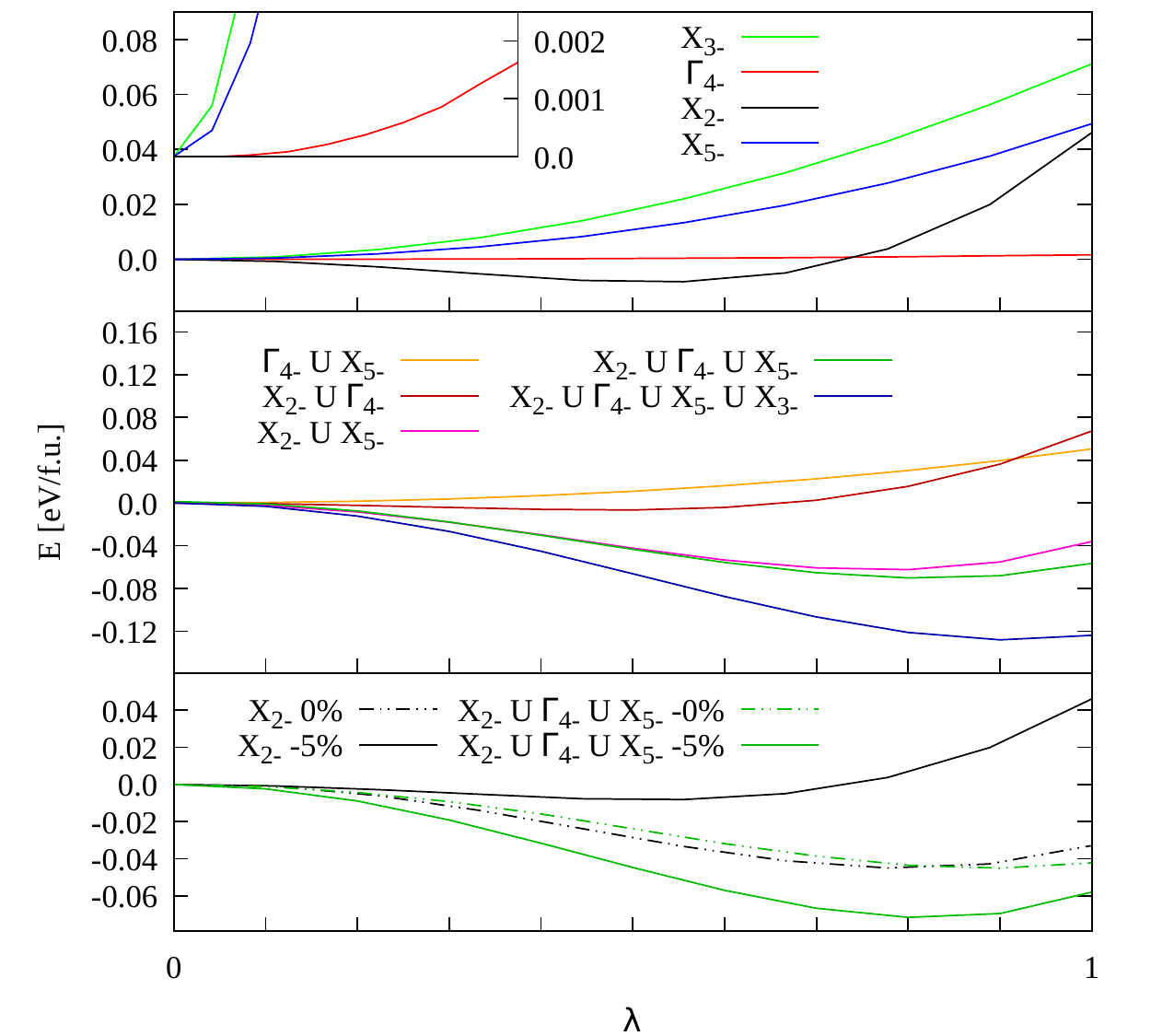}
}
\caption{ \label{modes}
 Symmetry modes analysis connecting the high temperature cubic phase to the ferroelectric R3m.
 The top panel reports the  {\it ab-initio} total energy for the dominant single-modes as a function of the generalized displacement coordinate $\lambda$.
 The central panel reports the energy trend of the coupled-dominant modes.
 The bottom panel reports the dependence of the leading order parameter X$_{2-}$ and of the coupling X$_{2-}$ U $\Gamma_{4-}$ U X$_{5-}$  on the generalized coordinate, in the case of 0\% and -5\% compressive strain.
 In the bottom panel the configurations at $\lambda=0$ are artificially set to zero to simplify the comparison: the -5\% undistorted phase is more than 220 meV/f.u. higher in energy with respect to the non-compressed phase.
}
\end{figure}
\begin{figure}[]
\centerline{
\includegraphics[angle=0,width=0.52\textwidth]{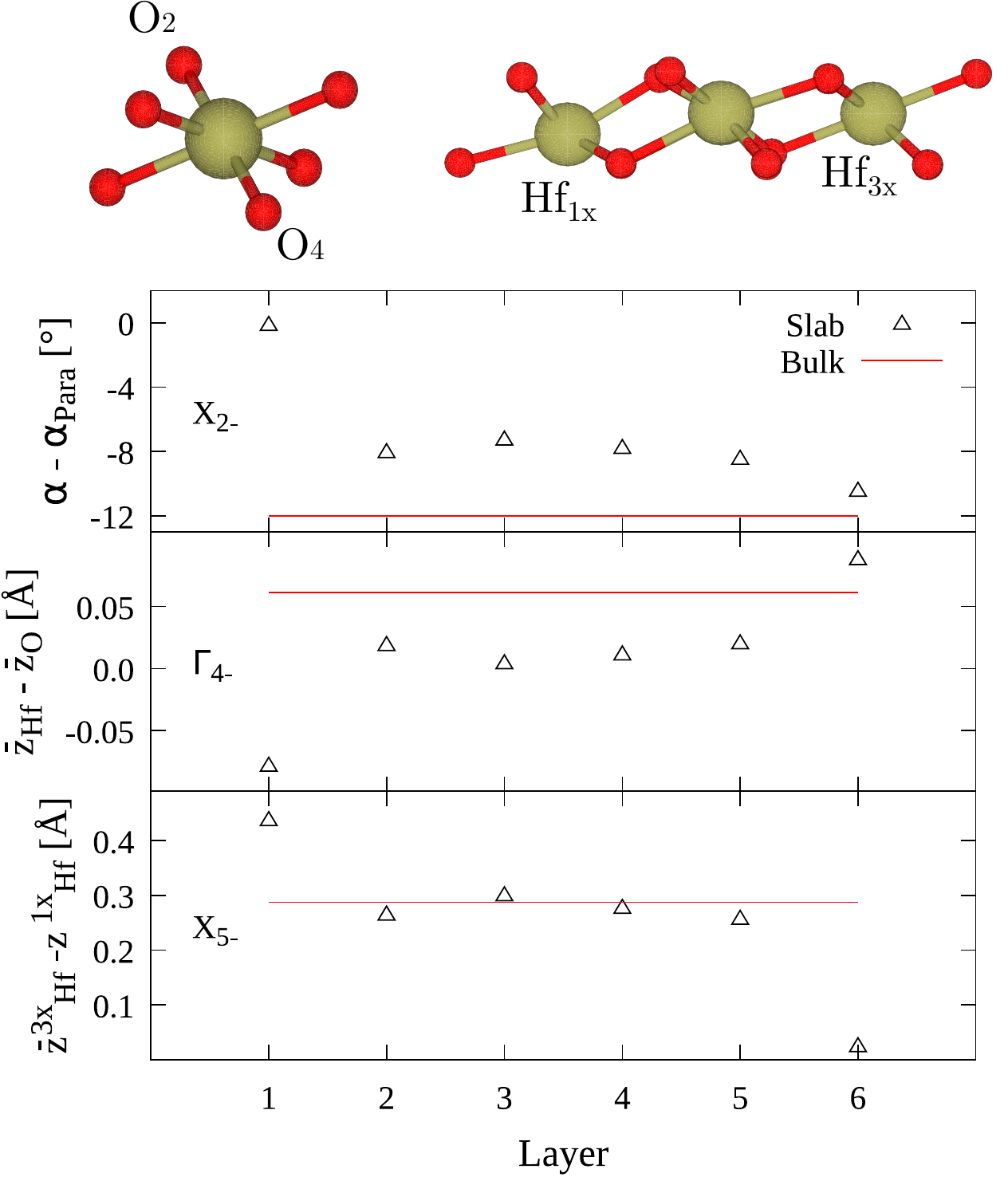}
}
\caption{ \label{inibition}
The evolution of the structural fingerprint for three dominant modes in the 6-layer slab.
The first panel reports the fingerprint for the leading order parameter. 
The angle $\alpha$ is the one connecting the O$_2$ and O$_4$ with the vertex at Hf in the structure above the panels at left.
The central panel reports the fingerprint for the polarization.
The panel at bottom reports the fingerprint for X$_{5-}$ mode.
It represents the z-component of the distance between Hf atoms with different Wyckoff positions which are reported in the picture above the panels at right.
}
\end{figure}
An alternative approach relies on the analysis of the main symmetry-allowed distortion modes connecting the hafnia  high-temperature cubic phase ($Fm-3m$) to the $R3m$ phase.
We employed standard crystallographic tools \cite{Stokes_2006,Orobengoa_2009} to identify the symmetry-adapted distortions and found six of them, two of which have negligible amplitudes. Five symmetry modes are associated with the wave vector q$_X$ at the Brillouin-zone boundary, and the last one to a polar distortion at $\Gamma$.
Under -5\% strain condition, the four dominant modes (in the hexagonal setting), ordered by decreasing amplitude, are: X$_{2-}$, X$_{5-}$,$\Gamma_{4-}$ and X$_{3-}$.
The top panel in Fig. \ref{modes} reports the {\it ab-initio} total energy computed along the distortions. 
The only soft mode is X$_{2-}$ corresponding to an instability of 8 meV/f.u. 
On the other hand, the high-symmetry phase is stable against all the remaining modes.
This is a canonical behaviour for an improper ferroelectric, where the polarization emerges through the coupling to the leading order parameter, which in this case is a zone-boundary distortion.
When examining the atomic distortions related to each mode, we note that the leading order mode distorts the angles formed by  a Hf atom with the surrounding oxygen cage and changes the relative bond lengths.
The X$_{5-}$ mode moves apart, along the \textbf{c} direction, the Hf with different Wyckoff positions. These correspond to the Hf atoms forming the primary and the secondary bonds in Fig.\ref{prsec_bonds}. 
The polar mode $\Gamma_{4-}$ moves apart the oxygen and the hafnium atoms along the \textbf{c} direction.
The X$_{3-}$ mode displaces the Hf forming the primary bonds along a non-trivial direction and those forming the secondary bonds simply along \textbf{c}.
Going back to Fig.\ref{modes}, the central panel reports the couplings between the different modes.
The coupling of the polarization with X$_{2-}$ does not significantly affect the soft character of the leading mode: the energy gain is 7 meV/f.u. . 
Instead, the X$_{5-}$ significantly enhances the soft character of the leading mode, the energy gain when coupled to the order parameter amounting to 54 meV/f.u. . 
It is worth noticing that including the $\Gamma_{4-}$ mode further stabilize this coupling, even though only by 9 meV/f.u. .
In this sense, we may see the coupling of the X$_{5-}$ mode with the order parameter as a condition favouring the excitation of the polar distortion.
When all the four dominant modes are coupled, the largest gain in energy is obtained: 128 meV/f.u. below the high symmetry phase at $\lambda = 0$.
The full relaxation, which includes the neglected structural distortions, leads to a
further energy decrease of 20 meV/f.u., {\em i.e.} less than one sixth of the energy gain corresponding to the simultaneous activation of the dominant modes.
The bottom panel in Fig.\ref{modes} describes the dependence of the energy dispersion of  X$_{2-}$ and of the coupling X$_{2-}$ U $\Gamma_{4-}$ U X$_{5-}$ on the applied strain.
The compression to -5\% weakens the character of the leading mode: its stability is reduced by almost 30 meV/f.u. . 
Instead, the three-modes-coupled distortion is strongly stabilized by the compression: when strain is applied, the energy gain is about 27 meV/f.u. .
We stress that this energetically favourable pairing with the compressive strain is not shown by the X$_{2-}$ U $\Gamma_{4-}$ distortion. This is a further sign of the special role assumed by X$_{5-}$ when compressive strain is applied.
Based on this analysis, we can describe the suppression of the polarization in the central portion of the slab in terms of suppression of the dominant modes. 
To this purpose, we define a structural ``fingerprint" for each mode, which consists in geometrical quantities catching unambiguously the distortions of each single mode.
From their trends throughout the slab, we can indeed deduce the variations of the modes amplitude.
The fingerprint describing the order parameter X$_{2-}$ can be taken as the variation of the angle $\alpha$ (between the oxygen labelled O$_2$ and O$_4$ in Fig.\ref{inibition} with the vertex located at the Hf atom), with respect to the reference value of $\pi$ ({\em i.e.} the $\alpha$ value in the PE phase at $\lambda=0$). 
This angle is also affected by $\Gamma_{4-}$, but the change deriving from this mode is much smaller (only  26\%) than the one from X$_{2-}$.
The other angles in the cage are significantly affected also by the mode X$_{3-}$: the distortion coming from this mode is almost 80\% of the one coming from X$_{2-}$. Therefore, they cannot be taken as fingerprints for the order parameter.
The polar distortion can be easily described in terms of the distance between the oxygens and the Hf average z-coordinates.
The effects of the other dominant modes on this quantity is negligible.
The X$_{5-}$ mode can be well represented by the finite distance between the z-coordinates of the Hf atoms having different Wyckoff positions. Specifically, we averaged the z-coordinate of the Hf with multiplicity 3.

Fig.\ref{inibition} reports the trend of the three structural fingerprints for the different layers in the 6-layer slab configuration. 
The first panel reports the difference of the  angle $\alpha-\alpha_{\rm para}$ representing X$_{2-}$ and its bulk value ($\approx$ 12°) as a function of the layer.
The reduction of about 30\% with respect to the bulk value, occurring in the central layers, is a clear indication of the mode suppression.
The central panel reports the trend for the fingerprint of the polar mode.
Also in this case we observe a clear reduction of the fingerprint in the central layers.
Given the detected improper character of ferroelectricity, we note that the suppression of the polarization should be interpreted as caused by the suppression of the X$_{2-}$ mode.
This dependence can also be inferred by the strong correlation between the trends of the two fingerprints: their suppression should not be intended as two independent phenomena.
The third panel illustrates that the X$_{5-}$ distortion is not suppressed as the depth increases.
On the contrary, its amplitude oscillates around the bulk value.
This is interesting to notice, given the considerable contribution to the stabilization of the polar phase coming from the coupling of this mode with the order parameter.
This suggests that  finite-size effects reduce the amplitude of the order parameter in such a way to lower the polarization in the central volume, while minimizing the reduction of energetic stability with respect to the pure bulk phase.
Given the connection with the polar mode  reported in Fig.\ref{modes}, this result also explains the incomplete suppression of the polarization. 
In fact, the finite size has no effect at all on the X$_{5-}$ mode, which is responsible for ``encouraging" the excitation of $\Gamma_{4-}$.
Therefore, the role of X$_{5-}$ when coupled with the order parameter is two-fold even in finite configuration: it contributes to the stabilization of the ferroelectric phase and it prevents the polar mode from completely vanishing.

\section{Conclusions}

In this paper we  addressed the effects of surfaces on the $R3m$ polar phase of hafnia.
We have shown how the introduction of the finite size leads to two types of consequences.
The first one consists in the contraction of the slab, resulting from the coupling between strain and the  order parameter.
This contraction leads to a reduction of  polarization by about one third compared to the bulk value. This effect should diminish with increasing thickness, but the thicknesses range that we analyzed does not allow us to observe this trend clearly. 
The second consequence is the relaxation of the atomic positions towards a configuration that further decreases the electric dipole density in the central layers. The values achieved are finite, but much smaller than the corresponding bulk values.
We explained these effects by means of two approaches: through an atomistic perspective, looking at bond lengths to identify polarization, and through a more fundamental approach based on symmetry-allowed distortion analysis. 
The trend in bond lengths throughout the slab clearly shows that the presence of the surfaces tends to balance the weight of different competing bonds that contribute with opposite sign to layer-polarization. This balance does not lead to full cancellation though, thus polarization is not completely suppressed.
The analysis of the distortions allowed us to identify the improper nature of  ferroelectricity in hafnium oxide rhombohedral phase, the leading-order parameter being the mode at the boundary-zone X$_{2-}$ instead of P. 
By assigning a structural fingerprint to each of the leading modes, we showed how the introduction of the surfaces leads to a suppression of the leading mode and, consequently, of the polarization in the central layers of the slab. However, even in this case the suppression is not complete and the amplitude of the distortions is not fully cancelled.
Finally, we found that these results are independent on the slab-thickness in the considered range (1-3.5 nm). Therefore, we may infer  the critical-thickness for the R3m phase of HfO$_2$ either to be  lower than 1 nm or to completely vanish.

\section{Acknowledgments}

We acknowledge support from the Italian Ministry for Research and Education through the PRIN-2017 project {\em ``TWEET: Towards Ferroelectricity in Two dimensions"} (IT-MIUR Grant No. 2017YCTB59) and computational support from CINECA through the ISCRA initiative with Grant N. HP10CEI2UQ (FeCoSMO project). We also acknowledge support from the computational resources at ``Gabriele d’Annunzio" University of Chieti.
\newline
\appendix

\section{Computational details}
\label{Methods}
We performed density functional theory simulations, as implemented in the Vienna Ab-initio Simulation Package (VASP) \cite{Kresse_1999} at the level of the revised Perdew, Burke, and Ernzerhof functional for solids \cite{Perdew_2008,Perdew_2009}. 
The bulk phases were relaxed until the difference in the total energy between two successive self-consistent steps is smaller than $10^{-7}$ eV and forces are smaller than $10^{-3}$ eV$\rm/\AA$.
The same convergence criteria apply for the 4-layers and 6-layers slabs.
Instead, the 12-layers slab was relaxed until all the forces are smaller than than $10^{-2}$ eV$\rm/\AA$.
The cutoff for the expansion on the plane waves basis is set to 600 eV. The k-point mesh is set to 5x5x3 for  bulk calculations and to 4x4x1 for the slabs.

We relax the slabs with at least 30 \AA\, of vacuum including dipole corrections to prevent periodic-images spurious interaction. No external field is applied. 

In the distortion modes analysis, we compute the energy of 10 intermediate structures along the distortion paths, with the same computational parameters reported above.
%


\end{document}